\documentclass[[aps,prc,twocolumn,showpacs,superscriptaddress]{revtex4}

\usepackage{bm}% bold math

\begin{document}

\title{Time Modulation of K-electron Capture Decay of Hydrogen-Like Ions with
    Multiphoton Resonance Transitions}

\author{I. M. Pavlichenkov}
 \email{pavi@mbslab.kiae.ru}
 \affiliation{Russian
Research Center "Kurchatov Institute", Moscow, 123182, Russia.}

\date{\today}% It is always \today, today,

\begin{abstract}
The multiphoton resonance transitions between ground hyperfine
states are used for the time modulation of the electron capture
decay of hydrogen like ions with the Gamow-Teller transition
$1^+\rightarrow 0^+$. The proposed mechanism offers a time
oscillating decay with the frequency up to 0.1 Hz. The experiment
to observe the modulation is proposed for ions stored in a Penning
trap. An attempt to understand the GSI anomaly with multiple
photon transitions is made.
\end{abstract}


\maketitle

Longstanding attempts have been made to change artificially the decay of
radioactive nuclei using different external factors. It was
suggested by Segr\'{e} \cite{Seg} that the orbital electron
capture (EC) decay rates depend on the density of atomic electrons
within a nucleus. Thus, environmental conditions may alter
electron densities and affect EC decay rates. However this effect
is small and does not exceed the fractions of percentage point.

The introduction of Penning traps and storage rings into nuclear
physics has allowed to study radioactive hydrogen-like (H-like)
ions. Recently the single ion spectroscopy technique with time
resolution less than 1 s has been developed at GSI Darmstadt. With
this elegant method, the oscillatory modulation of the exponential
EC decay for the $^{140}{\rm Pr}^{58+}$ and $^{142}{\rm Pm}^{60+}$
ions was found in \cite{Lit}. The modulation period of 7.1 s is
incommensurable with all the energy intervals involved in the
experiment with exclusion of the neutrino mass difference. Perhaps
this is the main motivation of the Letter \cite{Ivn} which relates
the anomaly to neutrino mixing.

Our motivation for this work is to show that there is a mechanism
related to Rabi oscillations which offers such a small period. To
observe the time modulation of the EC decay, we propose to apply
Penning trap mass spectrometry with ions in the intermediate mass
region having the "sterile" relative to the EC decay hyperfine
state. We analyze also the effect of the magnetic fields of dipole
and quadrupole magnets in the Experimental Storage Ring (ESR) on
the GSI oscillations.

The $1s_{1/2}$ state of a H-like ion has two hyperfine levels with
the total angular momentum $F=I\pm 1/2$, where $I$ is the nuclear
spin. In the case of a positive nuclear magnetic moment $\mu$, the
ground state of an ion has $F=I-1/2$, whereas for negative $\mu$,
the order of these energy levels is reversed. When $I=1$, the
hyperfine splitting of the $1s_{1/2}$ state is given by \cite{Sha}
\begin{equation}\label{spl}
\Delta E=\frac{2\alpha(\alpha Z)^3(\mu/\mu_N)
m^2c}{M(2\sqrt{1-(\alpha Z)^2}-1)\sqrt{1-(\alpha Z)^2} },
\end{equation}
where $m$, $M$, $\mu_N$, and $\alpha$ are the masses of electron
and proton, nuclear magneton, and the fine-structure constant,
respectively. For nuclei with $I^\pi=1^+$, the EC decay from the
$F=3/2$ state is forbidden because fully ionized daughter nuclei
have $I^\pi=0^+$. This feature of such ions may be used for
observing the time modulation of the EC decay.

The oscillatory behavior of the EC decay of such ions can be
obtained with the Rabi resonance method used in nuclear magnetic
spectroscopy. We consider an ion in a static magnetic field in the
$z$ direction, resulting in the Zeeman splitting of both hyperfine
levels. Transitions between the Zeeman components of these two
states are driven by the transverse oscillating field along the
$x$ direction with frequency $\omega$. When the frequency $\omega$
is tuned to $\Delta E/\hbar$, the oscillation of populations for
the states $F=1/2,M$ and $F=3/2,-M$, where $M=1/2$ or $-1/2$, is
modulated with the Rabi frequency, which is proportional to the
amplitude of the driving field. Because the EC decay probability
is proportional to the occupation of the $F=1/2$ state, the Rabi
frequency may be observed as the time modulation of the EC decay
rates. The time modulation of EC-decay is also possible with the
resonance transitions involving several field quanta.

{\it Penning trap.}--- A single H-like ion in a Penning trap is an
ideal object for observing the modulation of the EC decay. The ion
having a charge-to-mass ratio $q/M$ is confined in a strong
magnetic field ${\bf B}(0,0,B_0)$ superimposed with a weak
electrostatic quadrupole field \cite{Bro}. The solution of the
motion equations yields the cyclotron, axial and magnetron
harmonic oscillations. The ion mass is determined via the
determination of the free-space cyclotron frequency
$\omega_c=qB_z/Mc$ which derived from three measured harmonic
frequencies. The single ion spectroscopy developed in the ESR
experiments can be installed at a Penning trap \cite{Nov}. In this
technique, it is possible to realize the fully controllable
experiment on the time modulation of the EC decay by single- and
multi-photon resonance.

We are interested in the time evolution of the two hyperfine
states of an ion exposed by the trap magnetic field $\bf B$ and
irradiated by the in-plane elliptically polarized driving field
${\bf b}(b_x\sin\omega t,0,b_z\cos\omega t)$. The strength of the
latter is restricted by the frequency resolution of the single ion
spectroscopy. It is necessary for a shift in the cyclotron
frequency $\omega_c$ caused by the influence of the field ${\bf
b}$ to be small compared to the change in $\omega_c$ due to the
transition from a parent ion to a daughter nucleus. As a rough
approximation, this gives $b<B_0Q_{EC}/(Mc^2)$, where $Q_{EC}$ is
the decay energy.

The total Hamiltonian is $H=H_0+U(t)+V(t)$, where
$H_0=H_D+(e/2)[{\bm\alpha}{\bf r}]\!\cdot\!{\bf B}$ describes the
$1s_{1/2}$ state of a H-like ion in the Penning trap. Here $H_D$
is the Dirac Hamiltonian with hyperfine interaction and
$\bm\alpha$ is the Dirac matrix. The eigenfunctions and
eigenvalues of $H_0$ are $|+\rangle=|3/2,M_+\rangle$,
$E_{3/2}\!+\!(1/3)g\mu_BB_0M_+=\hbar\omega_3$ and
$|-\rangle=|1/2,M\rangle$,
$E_{1/2}\!-\!(1/3)g\mu_BB_0M=\hbar\omega_1$. Here $\mu_B$ is the
Bohr magneton and $g$ is the gyromagnetic ratio of the bound
electron in a H-like ion. The terms
$U(t)\!=\!(e/2)[{\bm\alpha}{\bf r}]_zb_z\cos\omega t$ and
$V(t)\!=\!(e/2)[{\bm\alpha}{\bf r}]_xb_x\sin\omega t$ represent
the diagonal and off-diagonal parts of the interaction between an
ion and the driving field.

There are three independent transitions between the lower and
upper Zeeman sub-states which can be used for the modulation of the
EC decay: \\
(i) The transitions between the states $|3/2,+1/2\rangle$ and
$|1/2,-1/2\rangle$ or $|3/2,-1/2\rangle$ and $|1/2,+1/2\rangle$
with the energy $\Delta
E\!=\!E_{3/2}\!-\!E_{1/2}\!=\!\hbar\omega_0$
regardless of the sign of $\mu$.\\
(ii) The transitions between the states $|3/2,\mp 3/2\rangle$ and
$|1/2,\mp 1/2\rangle$ for $\mu>0$ (upper sign) and $\mu<0$ (lower
sign) with the energy $\Delta E-(2/3)g\mu_BB_0$.\\
(iii) The transitions between the states $|3/2,\pm 3/2\rangle$ and
$|1/2,\pm 1/2\rangle$ for $\mu>0$ (upper sign) and $\mu<0$ (lower
sign) with the energy $\Delta E+(2/3)g\mu_BB_0$.\\
Thus, we have the two discrete states $|+\rangle$ and $|-\rangle$
with the energies $\epsilon_3\!=\!\hbar\omega_3$ and
$\epsilon_1\!=\!\hbar(\omega_1-i\lambda/2)$, where $\lambda$ is
the EC decay constant. Let the amplitudes for an ion to be in
these states are $a_+(t)$ and $a_-(t)$. Then the system evolves
according to the equations
\begin{eqnarray}
i\hbar\dot{a}_-&\!\!=\!\!&(\epsilon_1+
U_-\cos\omega t)a_-\!+V_{-+}a_+\sin\omega t,\nonumber\\
i\hbar\dot{a}_+&\!\!=\!\!&(\epsilon_3+ U_+\cos\omega t)a_+
+V_{+-}a_-\sin\omega t, \label{diab}
\end{eqnarray}
where $U_+\!\!=\!\!=g\mu_Bb_zM_+/3$,
$U_-\!\!=\!\!=-g\mu_Bb_zM_-/3$,
\begin{equation}
V_{+-}\!\!=\!\!\left\{\begin{array}{l}\hbar V=
\frac{1}{3\surd{2}}g\mu_Bb_x\quad \mbox{\small{for
(i),}}\\
\hbar\tilde{V}=-\frac{M_+}{\surd{3}|M_+|}g\mu_Bb_x\quad
\mbox{\small{for (ii) and (iii)}}.
\end{array}\right.
\end{equation}

To eliminate the diagonal oscillating terms in (\ref{diab}), we
perform the transformation
\begin{equation}
a_\pm=A_\pm\exp{\left[-i(\omega_1+\omega_3)t/2-
i(U_\pm/\hbar\omega)\sin{\omega t}\right]},
\end{equation}
which reduces the equations for the transitions (i) to the form
\begin{eqnarray}
i\dot{A}_-\!\!&=&\!\!-\textstyle{\frac{1}{2}}(\omega_0+i\lambda)A_-
+VA_+\sin\omega t,\nonumber\\
i\dot{A}_+\!\!&=&\!\!\textstyle{\frac{1}{2}}\omega_0A_+
+VA_-\sin\omega t, \label{diab0}
\end{eqnarray}
The two other transitions are described by the equations
\begin{eqnarray}
i\dot{A}_-\!\!&=&\!\!\!-\textstyle{\frac{1}{2}}(\tilde{\omega}_0\!+\!i\lambda)A_-
\!+\!\tilde{V}A_+\exp{\left(i\textstyle{\frac{U}
{\omega}}\sin\omega t\right)}\sin\omega t,\nonumber\\
i\dot{A}_+\!\!&=&\!\!\textstyle{\frac{1}{2}}\tilde{\omega}_0A_++
\tilde{V}A_-\exp{\left(-i\textstyle{\frac{U} {\omega}}\sin\omega
t\right)}\sin\omega t,
\end{eqnarray}
where $\tilde{\omega}_0=\omega_3-\omega_1$, $U=(U_--U_+)/\hbar$.
Through the use of the expansion of an exponent in the Bessel
functions, these equations are converted to
\begin{eqnarray}
i\dot{A}_-\!\!&=&\!\!-\textstyle{\frac{1}{2}}(\tilde{\omega}_0\!+\!i\lambda)A_-
-i\tilde{V}A_+\hspace{-2mm}\sum\limits_{m=-\infty}^\infty\hspace{-1mm}
J'_m\left(\frac{U}{\omega}\right)e^{im\omega t},\nonumber\\
i\dot{A}_+\!\!&=&\!\!\textstyle{\frac{1}{2}}\tilde{\omega}_0A_+
+i\tilde{V}A_-\hspace{-2mm}\sum\limits_{m=-\infty}^\infty\hspace{-1mm}
J'_m\left(\frac{U}{\omega}\right)e^{-im\omega t}, \label{diab12}
\end{eqnarray}
where $J'_m$ is the derivative of the Bessel function. The
equations (\ref{diab0}) and (\ref{diab12}) describe an ion with a
positive nuclear magnetic moment $\mu$. When $\mu<0$, it is
necessary to change the sign of $\omega_0$ and $\tilde{\omega}_0$.
Up to this point all of the equations are exact.

The equations (\ref{diab0}) for $\lambda=0$ are the basic ones in
the theory of the resonance interaction of the two level system with
a periodic field. When $\omega\approx\omega_0$, we get the usual
Rabi formula for amplitudes replacing the sine by one exponent
\cite{Lan}. For the K-photon resonance when $\omega_0\approx
K\omega$, Shirley \cite{Shi} found the approximate solution for a
weak driving field. In the adiabatic approximation, $K\gg 1$, the
solution of (\ref{diab0}) was first obtained by Zaretskii and
Krainov \cite{Zar} and then by Duvall {\it et al.} \cite{Duv}. In
the weak field regime, $b_x\ll B_0$, which is adequate for the
Penning trap experiment, we use Sambe's perturbation approach
\cite{Sam} provided that $\lambda\ll\omega$. In the $K$-th order
of the perturbation theory in the quasienergy representation, Eqs.
(\ref{diab0}) can be approximated by
\begin{eqnarray}
i\dot{A}_-\!\!&=&\!\!-\textstyle{\frac{1}{2}}(\omega_{res}-
K\omega+i\lambda)A_-+\textstyle{\frac{1}{2}}\Gamma_KA_+,\nonumber\\
i\dot{A}_+\!\!&=&\!\!\textstyle{\frac{1}{2}}(\omega_{res}
-K\omega)A_++\textstyle{\frac{1}{2}}\Gamma_KA_-, \label{reduc}
\end{eqnarray}
where \vspace{-4mm}
\begin{equation}
\omega_{res}=\omega_0\left[1+\frac{K^2}{K^2-1}
\left(\frac{g\mu_Bb_x}{3\sqrt{2}\hbar\omega_0}\right)^2\right],
\quad K>1
\end{equation}
is the shift of the resonance frequency related to the dynamic
Zeeman effect which is referred as the Bloch-Siegert shift. The
shift is of the order of $\omega_0(b_x/B_0)$. For $K=1$ the
coefficient before round brackets is equal to $1/4$. The half
width of $K$-photon resonance is  \vspace{-2mm}
\begin{equation}\label{hwl}
\Gamma_K=\frac{2\omega}{[(K-1)!!]^2}
\left(\frac{g\mu_Bb_x}{6\sqrt{2}\hbar\omega}\right)^K,
\end{equation}
the photon number $K$ being odd for these transitions.

We now turn to a close examination of Eqs. (\ref{diab12}) for the
transitions (ii) and (iii). When the frequency $K\omega$ ($K$ is
even or odd integer) is nearly resonant with the energy separation
$\tilde{\omega}_0$ of the two states, the term with $m=K$
represent a resonance driving field, whereas those with $m\neq K$
result in the resonance frequency being shifted. In the second
order perturbation theory, we obtain
\begin{equation}\label{ore}
\omega_{res}\!=\!\tilde{\omega}_0\!\left\{\!\!1\!+\!
2\!\left(\frac{g\mu_Bb_x}{\sqrt{3}\hbar\tilde{\omega}_0}\right)^2
\!\!\!\!\sum_{m=-\infty\atop m\neq K}^\infty\!
\!\!\!\frac{\left[J'_m\!\left(\frac{2g\mu_Bb_z}
{3\hbar\omega}\right)\right]^2}{1-\frac{m}{K}}\!\right\}\!.
\end{equation}
Substitution of $A_\pm$ by $A_\pm\exp[\mp i(K\omega t-\pi/2)/2]$
leads us to (\ref{reduc}) with $\omega_{res}$ given by (\ref{ore})
and the half width
\begin{equation}\label{hwc}
\Gamma_K=\frac{2g\mu_Bb_x}{\sqrt{3}\hbar}\left|J'_K\!
\left(\frac{2g\mu_Bb_z} {3\hbar\omega}\right)\right|.
\end{equation}

We assume that an ion is in the lower hyperfine state when it is
injected in a Penning trap at $t=0$. With this initial condition
Eqs. (\ref{reduc}) can be easily solved for $\lambda\ll\Gamma_K$.
The probability of finding an ion in the decaying state
$|-\rangle$ is given by \vspace{-1mm}
$$
P_F(t)=\delta_{FF_-}
\left[\cosh\left(\frac{\lambda\Delta}{2\Omega}t\right)-
\frac{\Delta}{\Omega}\sinh\left(\frac{\lambda\Delta}{2\Omega}t\right)
\right]e^{-\lambda t/2}+\nonumber
$$
\vspace{-5mm}
\begin{equation}
\frac{(\delta_{FF_+}\!-\!\delta_{FF_-})
\Gamma^2_{res}}{\Delta^2+\Gamma^2_{res}}\left[
\sinh^2\left(\frac{\lambda\Delta}{4\Omega}t\right)\!
+\!\sin^2\left(\frac{1}{2}\Omega t\right)\right]e^{-\lambda t/2},
\label{pro}
\end{equation}
where $\Omega=\sqrt{\Delta^2+\Gamma^2_K}$ is the Rabi frequency
and $\Delta=\omega_{res}- K\omega$ is the detuning. The
probability depends on the lower state angular momentum $F=F_\pm$.
Expression (\ref{pro}) describes damped oscillations with the
frequency $\Omega$ decaying on the scale of {\it half} the EC
decay constant $\lambda$. Nevertheless, the total number of decays
for infinite time calculated with (\ref{pro}) is equal to the
number of radioactive ions at $t=0$.

To elucidate the time modulation of EC decay, we utilize the
concept of the dressed atom introduced by Shirley \cite{Shi} as
the quantum limit of the semiclassical theory. The ion with
$\mu>0$ coasted in a trap and exposed to a resonance field is
described by the quantum wave function
\begin{equation}\label{ent}
\Psi(t)\!=\!a_+(t)|+\rangle|N\!-\!K\rangle+
a_-(t)|-\rangle|N\rangle,
\end{equation}
where $|N\rangle$ and $|N\!-\!K\rangle$ are the wave functions
with the fixed number of photons having the energy $\hbar\omega$.
Thus, the dressed ion has the two hyperfine states entangled with
photons. The EC decay at the time $t$ disentangles the state
(\ref{ent}) by projecting $\Psi(t)$ on the state $|-\rangle$ with
the probability $P_{F_-}(t)$ (\ref{pro}), to which the decay rate
is proportional.

To observe the time modulation, the H-like radioactive
ion has to have the following properties: \\
(1) Its half-life $T_{1/2}$ should be longer than its
preparation and cooling time. \\
(2) It should have a large fraction of the EC decay, implying large $Z$. \\
(3) The life time $\tau$ of the upper hyperfine state should be
longer than $T_{1/2}$, implying small $Z$.  \\
(4) The frequency $\Omega$ of the time modulation of the EC decay
should be greater than the EC decay constant $\lambda$. \\
Table~\ref{tab:table1} shows H-like ions which may be considered
as candidates for the experimental observation of time modulation.
The hyperfine splitting $\Delta E$ is calculated by Eq.
(\ref{spl}). The mean time $\tau$ for the magnetic dipole
transition between the upper and lower hyperfine states is
estimated with the expression \cite{Sob}\vspace{-2mm}
\begin{equation}
\tau=\frac{3\lambda_e}{2\alpha c}\left(\frac{mc^2}{\Delta E}
\right)^3\!\frac{2I+1}{2F+1},
\end{equation}
where $\lambda_e$ is the electron Compton wavelength and $F$ is
the angular momentum of the ground state. The H-like ion
$^{68}{\rm Ga}$ seems to be the best candidate. In
Table~\ref{tab:table2} we have compared the amplitudes of the two
fields with different polarizations calculated from Eqs.
(\ref{hwl}) and (\ref{hwc}) for different photon number $K$ and
the fixed width $\Gamma_K$. The trap field is $B_0=1.0$ T. It is
seen that the circularly polarized field is more suitable for
experimental use.
\begin{table}[t]
\caption{\label{tab:table1}Hydrogen-like radioactive ions with
allowed Gamow-Teller transition $1^+\rightarrow 0^+$ appropriate
for observation of the time modulation of the EC decay. The
half-lives $T_{1/2}$ of nuclear ground state is given for neutral
atoms.}
\begin{ruledtabular}
\begin{tabular}{cccccccc}
 H-like&$\mu$&$\Delta E$&$\tau$& $T_{1/2}$&$Q_{EC}$&EC/$\beta^+$ \\
  ion  &$\mu_N$ &meV       &      &          &keV&      \%       \\
\hline $^{18}{\rm F}$&+0.58\footnotemark[1] &0.672 & 5.55y &109.8m&1655&3.3/96.7 \\
$^{64}{\rm Cu}$&$-0.217\footnotemark[2]$&8.95&10.3h&12.7h&1673&43/18  \\
$^{68}{\rm Ga}$&0.0117\footnotemark[2]&0.596&7.95y &67.6m&2921&8.9/88.0\\
$^{78}{\rm Br}$&0.13\footnotemark[2]&9.75&15.9h&6.46m&3754&8/92\\
$^{82}{\rm Rb}$&+0.554\footnotemark[2]&48.0&7.20m&1.27m&4378&4/96\\
\end{tabular}
\end{ruledtabular}
\footnotetext[1]{Calculated with Schmidt's factors $g^p_j$ and
$g^n_j$.} \footnotetext[2]{The data from Ref. \cite{Sto}. No sign
is given, if it is unknown.}
\end{table}
\begin{table}[h]
\vspace{-3mm} \caption{\label{tab:table2}Strength of the linearly
polarized ($b_l=b_x$ for transition (i) with $\omega_0/2\pi=144$
GHz) and circularly polarized ($b_c=b_x=b_z$ for transition (ii)
with $\tilde{\omega}_0/2\pi=126$ GHz) harmonic fields required to
observe the $K$-photon resonance with the half width
$\Gamma_K/2\pi=0.1$ Hz in the ion $^{68}{\rm Ga}^{30+}$.}
\begin{ruledtabular}
\begin{tabular}{cccccc}
$K$&1&2&3&4&5\\
\hline
$b_l$,T&$1.54\!\cdot\!10^{-11}$&$-$&$2.38\!\cdot\!10^{-3}$&$-$&$9.07\!\cdot\!10^{-2}$\\
$b_c$,T&$1.00\!\cdot\!10^{-12}$&$2.62\!\cdot\!10^{-6}$&
$3.47\!\cdot\!10^{-4}$&$3.94\!\cdot\!10^{-3}$&$1.69\!\cdot\!10^{-2}$\\
\end{tabular}
\end{ruledtabular}
\end{table}

{\it Storage ring ESR.}---We try now to use multiple transition
mechanism for the interpretation of the GSI oscillations. It
should be emphasized, however, that this problem is a more
complicated one owing to the extremely complex configuration of
the ring magnetic field designed to store ions. Moreover, the
available experimental information is yet incomplete.

The ESR lattice has periodic structure with the spatial period
$L=C/2$, where $C=108.36$ m is the circumference of the ring.
Thus, an ion circulating in the ESR is exposed to the periodic
magnetic fields of bending dipole and quadrupole magnets. We
consider the two fields, which are in the direction of the x-axis
tangent to the main orbit and the z-axis perpendicular to its
plane. In the reference frame of a moving ion they can be
expressed  by the Fourier series \vspace{-2mm}
\begin{eqnarray}
b_x(t)&=&\sum^\infty_{n=1}b_{xn}\sin\left(2\pi n\gamma
vt/L\right), \nonumber \\
B_z(t)&=&B_{z0}+\sum^\infty_{n=1} b_{zn} \cos\left(2\pi n\gamma
vt/L\right), \label{fur}
\end{eqnarray}
where $t$ is the proper time. For the ions in consideration the
Lorentz factor is $\gamma=1.43$. The frequency $\omega_n$ of
$n$-th harmonic is equal to $n\omega$, where $\omega/2\pi=\gamma
v/L=5.65$ MHz is the frequency of the first harmonic and the
interval between nearby harmonics. The latter is much greater that
the width of the multiphoton resonance at the frequency
$\omega_n$. Therefore, the resonances with different $n$ do not
overlap, and may be considered as independent.

At first we consider the transition, induced by the fields of
$n$-th harmonic, between the hyperfine states $F=1/2$ and $F=3/2$
with the energy $\Delta E\sim 1$ eV. In principle, the resonance
transitions with the Rabi frequency of 0.1 Hz is possible for very
high $n$ and small $K$. However, this is a fallacious result,
because the mean life time of the upper $F=3/2$ state, $\tau=0,03$
s, is much shorter than the modulation period. Hence, the
"sterile" state can not be the source of observed oscillations.

Let us next consider the transitions between the Zeeman components
of the lower hyperfine state. For the modulation to be possible,
the weak interaction operator of the EC decay has to have the
different matrix elements for the states with different $M$. We
assume that this is the case. Then the two quasi-stationary states
exposed to the fields of n-th harmonic evolve according to
(\ref{diab}), where $a_+$ and $a_-$ are the amplitudes of the
lower $|1/2,+1/2\rangle$ and upper $|1/2,-1/2\rangle$ states with
energies $\epsilon_1=\hbar(\omega_+-i\lambda/2)$ and
$\epsilon_3=\hbar(\omega_--i\lambda/2)$, respectively.

Suppose the ion $^{140}{\rm Pr}^{58+}$ (or $^{142}{\rm Pm}^{60+}$)
circulating in the ESR moves along the main orbit. Then it is
exposed to the periodic magnetic field of dipole magnets only. One
period $L$ contains three identical dipoles arranged symmetrically
around the center of the central dipole magnet. The dipole magnet
of length $d=6.5$ m involves the strong bending field ${\bf
B}(0,0,B_0)$, $B_0=1,197$ T, and the fringe field ${\bf
b}(b_x,b_y,b_z)$. The latter is calculated by the code OPERA
\cite{Dol}. It is localized in the small region $\Delta x=29$ cm
on the two sides of each dipole. The component $b_z$ decreases
monotonically from $B_0$ to zero. The component $b_x$ has a pulse
shaped form with the maximal value $b_0=36$ mT, and the component
$b_y$ is too small to be taken into account. The Fourier
coefficients in (\ref{fur}) are easily found for these components
and the matrix elements in (\ref{diab}) are $U_+=U_-=pf_n/n,
V_{+-}=V_{-+}=-qg_n/n$, where
\begin{equation}\label{dat}
\ p\!=\!\frac{g\mu_B\gamma B_0}{\pi},\ q\!=\!\frac{g\mu_B
b_0}{\pi},\ {\rm and}\ \ \varpi\!=\!\frac{g\mu_B\gamma B_0d}{\hbar
L_0}
\end{equation}
is the Zeeman splitting $\varpi=\omega_--\omega_+$. The periodic
in $n$ functions $f_n\sim g_n\lesssim 1$ depend on the lattice
parameters. The  splitting $\varpi/2\pi=5.65$ GHz is well above
the frequency of the first harmonic. This means that the harmonics
with high $n$ are involved in the resonance transitions. Hence,
the driving field is weak and we can use the approximate solution
of (\ref{diab12}). In Table III we show the half width and
detuning for the $K$-photon resonance calculated according to the
formulas
\begin{equation}
\Gamma_K=\left|\frac{qg_n}{\pi\hbar n}
J'_{K}\left(\frac{2pf_n}{n\hbar\omega_n}\right)\right|,\
\Delta=\varpi-K\omega_n, \label{widt}
\end{equation}
for the harmonics, which are closest to the resonance.
\begin{table}[h]
\caption{\label{tab}Half width and detuning for the harmonics
close to the $K$-photon resonance for the ion $^{140}{\rm
Pr}^{58+}$ or $^{142}{\rm Pm}^{60+}$ moving along the main orbit.
Here $K$ is integer value of $\varpi/\omega_n$. All frequencies
are in the laboratory frame.}
\begin{ruledtabular}
\begin{tabular}{cccccccc}
 $n$&$K$&$\Gamma_K$\hspace{-0.5mm} (Hz)&$\Delta$\hspace{-0.5mm} (MHz)&
 $n$&$K$&$\Gamma_K$\hspace{-0.5mm} (Hz)&$\Delta$\hspace{-0.5mm} (MHz)\\
\hline 143& 7 &$1.48\cdot 10^{-10}$ & $-6.92$ &250
& 4 & 24.4 &$-2.97$ \\
166& 6 &$2.88\cdot 10^{-5}$& 12.9 &251
& 4 & 0.304 & $-18.8$ \\
167& 6 &$5.22\cdot 10^{-4}$&$-10.9$ &333
& 3 & 18.9 & 0.989 \\
168& 6 & 0.328 &$-34.6$&500
& 2 & 854 &$-2.79$ \\
201& 5 &0.782& $-22.7$ &999
& 1 &2420& $0.989$  \\
\end{tabular}
\end{ruledtabular}
\end{table}

The large values of $\Delta$ in Table~\ref{tab} may be attributed
to the inaccurate trajectory being used. In a storage ring,
periodic motion of an ion in the axial direction (rotation) is
coupled with its periodic motion in the transverse direction
(betatron oscillations). The latter leads to an ion being exposed
to the magnetic field of quadrupole lenses which change $\Gamma_K$
and $\Delta$. However, for the resonance condition $\Delta\sim
\Gamma_K$ to be realized an ion has to move along a particular,
"resonance" orbit. The decays of ions stored on non-resonance
orbits generate the background with the conventional exponential
law.

The plausible explanation of the observed oscillation amplitude is
related to the coupling of the $F=1/2$ and $F=3/2$ states. The
Hamiltonian corresponding to Eqs. (\ref{diab12}) offers the
coupling due to virtual transitions in the states F=1/2 and 3/2
with different $M$ and nonzero photon numbers. The transitions
change the energy (like the Lamb shift) and the wave functions of
the states $|1/2,\pm 1/2\rangle$. However, these changes are small
due to the weak driving field of the n-th harmonic and cannot
explain the observed amplitude.

It is not inconceivable also that the coupling of these states is
effected by the static or periodic potential $W$ associated with
the ESR fields. The solution of the Schr\"{o}diger equation for
the two hyperfine states gives in the limit $W^{1/2,\pm
1/2}_{3/2,M}/\Delta E\ll 1$ (the matrix elements $W$ is assumed to
be real) the new eigenfunctions
\begin{equation}\label{basis}
\left[\begin{array}{c}\!\!|1\rangle\!\!\\\!\!|2\rangle\!\!\end{array}\right]\!=
\!\!\left(\begin{array}{cc}\cos{\frac{\theta}{2}}
&\!\sin{\frac{\theta}{2}}\!\\\!-\sin{\frac{\theta}{2}}\!
&\cos{\frac{\theta}{2}}\end{array}\right)\!\!
\left[\begin{array}{c}\!|\frac{1}{2},\frac{1}{2}\rangle\!\!\\
\!|\frac{1}{2},-\frac{1}{2}\rangle\!\!\end{array}\right]\!,
\tan\theta\!=\!\frac{2(W\!\!\cdot\!\!
W)^{\frac{1}{2},\frac{1}{2}}_{\frac{1}{2},-\frac{1}{2}}}
{\hbar\varpi\Delta E}.
\end{equation}
With this basis, we find, following the method of \cite{Sam}, the
quasienergy function
\begin{equation}\label{func}
\Psi(t)\!=\!\left(b_1|1\rangle + b_2|2\rangle
e^{-iK\omega_nt}\right)e^{-\frac{i}{2}
(\omega_-+\omega_+-i\lambda)t},
\end{equation}
satisfying the initial condition $\Psi(0)=|1\rangle$. Here the
coefficients
$$
b_1\!=\!\cos\left(\Omega t/2\right)\!
-\!i\frac{\Gamma_K}{\Omega}\sin\left(\Omega t/2\right),\
b_2\!=\!-i\frac{\Delta}{\Omega}\sin\left(\Omega t/2\right)
$$
represent Rabi oscillations (slow motion). The function
(\ref{func}) describes the evolution of the system and allows to
calculate the amplitude of the transition of the mother ion from
the state $1s_{1/2}, F=1/2$ into the daughter nucleus. The
probability of the EC decay is proportional to the square of the
amplitude averaged over the period of n-th harmonic (fast motion)
\begin{equation}
P_{\!EC}\!\sim\!|M|^2\!\left[\!1\!-\!\sin\theta+
\frac{2\Gamma^2_K}{\Omega^2}\sin\theta
\sin^2\left(\frac{\Omega}{2}t\right) \right]\!e^{-\lambda t}\!,
\label{prob}
\end{equation}
where $|M|^2=|M_{+\frac{1}{2}}|^2=|M_{-\frac{1}{2}}|^2$ are the
squared amplitudes of EC decay calculated with the unperturbed
states. Equation (\ref{prob}) describes damped Rabi oscillations
with the EC decay constant $\lambda=\lambda_{EC}$ and the
amplitude $\sin\theta\sim W^2/(\hbar\varpi\Delta E)$. Comparing
the latter with observed amplitude 0.2, we find $W\sim0,001$ eV.
The electric quadrupole field required to achieve this value is of
the order of atomic fields and cannot exist in the ESR. The
required magnetic field ${\bf B}(0,B_y,B_z)$ should be of the
order of 20 T, which is greater by a factor 20 than the ESR
fields.

{\it Conclusion.}---The effect of atomic electron on the EC decay
of H-like ions is very important. We have shown for the first time
that the EC-decay rates of such ions can be modulated by using the
Rabi resonance method with single or multiphoton transitions. For
a long time the multiphoton resonance has been subject of intense
experimental and theoretical studies in different fields except
nuclear physics. We have proposed the experiment with intermediate
mass ions stored in a Penning trap which can demonstrate this
phenomenon.

In an attempt to understand the GSI oscillations, we have proposed
an alternative mechanism involving resonance multiphoton
transitions, induced by the periodic magnetic field of the ESR,
between the magnetic sub-states of the ground $F=1/2$ state of the
radioactive ions $^{140}{\rm Pr}^{58+}$ and $^{142}{\rm
Pm}^{60+}$. The mechanism includes the coupling of $F=1/2$ and
$F=3/2$ hyperfine states to explain the oscillation amplitude. The
important point of this scenario is that the predicted EC decay
constant does not violate experimentally and theoretically
established ratio
$\lambda^{H-like}_{EC}/\lambda^{He-like}_{EC}\approx 3/2$
\cite{Winc}. We find that the periodic magnetic field of the ESR
can generate the oscillations with the frequency of about 0.1 Hz
in the transitions involving several photons. However, the field
required to explain the observed amplitude should be many times
higher than that of the ESR. Moreover, to tune to the resonance a
stored ion should move along specific, "resonance" orbits. This
cast some doubts upon the involvement of the hyperfine structure
in the observed oscillations. Therefore, it is necessary to
perform the single-ion experiment with the ions $^{123,5}{\rm
Cs}^{54+}$ having the ground state $1s_{1/2}F=0$ or with the
He-like ions without a hyperfine structure. From these experiments
definite conclusions may be drawn regarding competing hypotheses
for the GSI oscillations.

{\it Acknowledgments.}--- The author is grateful to  A. Dolinski
for helpful information concerning storage ring physics, and to A.
L. Barabanov and S. T. Belyaev for critical reading of the
manuscript and useful comments. The work was supported by the
Grant NS-3004.2008.2.


\begin{thebibliography}{99}
\bibitem{Seg}E. Segr\'{e}, Phys. Rev.\ {\bf 71}, 274 (1947).
\bibitem{Lit} Y. Litvinov {\it et al.}., Phys.\ Lett.\ B {\bf 664},
162 (2008).
\bibitem{Ivn}A. N. Ivanov and P. Kienle, Phys.\ Rev.\ Lett.\ {\bf 103},
062502 (2009).
\bibitem{Sha}V. M. Shabaev {\it et al.}, Phys. Rev.\ A\ {\bf 37}, 4685 (1988).
\bibitem{Bro}L. S. Brown and G. Gabrielse, Rev.\ Mod.\ Phys.\ {\bf 58},
233 (1986).
\bibitem{Nov}Yu. Novikov, private communication (2009).
\bibitem{Lan}L. D. Landau and E. M. Lifshitz // {\it Quantum Mechanics}
(Pergamon, Oxford, 1958).
\bibitem{Shi}J. H. Shirley, Phys. Rev.\ {\bf 138}, B979 (1965).
\bibitem{Zar}D. F. Zaretskii and V. P. Krainov, Zh. Eksp. Teor. Fiz.
{\bf 66}, 537 (1974) [Sov. Phys. JETP. {\bf 39}, 257 (1974)].
\bibitem{Duv}R. E. Duvall, E. J. Valeo, and C.R. Oberman,
Phys. Rev.\ A\ {\bf 37}, 4685 (1988).
\bibitem{Sam}H. Sambe, Phys. Rev.\ A\ {\bf 7}, 2203 (1973).
\bibitem{Sob}I. I. Sobelman// {\it Atomic Spectra and Radiative Transitions}
(Springer-Verlag, Berlin/New York, 1979).
\bibitem{Sto}N. J. Stone, At. Data Nucl. Data Tables\ {\bf 90}, 75 (2005).
\bibitem{Dol}A. Dolinski, private communication (2009).
\bibitem{Winc}N. Winckler {\it et al.,
arXiv:0907.2277v1[nucl-th]}.
\end{thebibliography}
\end{document}